\documentclass[preprint2]{aastex}

\usepackage{multicol}
\usepackage{natbib}
\usepackage{amsmath}
\slugcomment{Not to appear in Nonlearned J., 45.}

\shorttitle{The turbulent evolution of the ISM at different metallicities.}
\shortauthors{Walch et al.}

\begin{document}


\title{The turbulent fragmentation of the interstellar medium:
 The impact of metallicity on global star formation.}
% and its connection to the boring fate of dwarf galaxies / effect on star formation.}


\author{S. Walch\altaffilmark{1}}
\email{stefanie.walch@astro.cf.ac.uk}
%\affil{}

\author{R. W\"unsch\altaffilmark{2}}
\email{richard@wunsch.cz}

\author{A. Burkert\altaffilmark{3}}
%\email{andi@usm.lmu.de}

\author{S. Glover\altaffilmark{4}}
%\email{s.glover}

\and

\author{A. Whitworth\altaffilmark{1}}
%\affil{School of Physics and Astronomy, Cardiff University, Queens Buildings, 5 The Parade, CF24 3AA, Cardiff, UK}
%\email{Anthony.Whitworth@astro.cf.ac.uk}


\altaffiltext{1}{School of Physics and Astronomy, Cardiff University, Queens Buildings, 5 The Parade, CF24 3AA, Cardiff, UK}
\altaffiltext{2}{Astronomical Institute, Academy of Sciences of the Czech Republic, Bocni II 1401, 141 31 Prague, Czech Republic}
\altaffiltext{3}{Universit\"ats-Sternwarte M\"unchen, Department f\"ur Physik, LMU M\"unchen, Scheinerstr.~1, 81679 Munich, Germany}
\altaffiltext{4}{Zentrum f\"ur Astronomie der Universit\"at Heidelberg, Institut f\"ur Theoretische 
Astrophysik, Albert-Ueberle-Str. 2, 69120 Heidelberg, Germany}


\begin{abstract}
We study the influence of gas metallicity, turbulence, and non-equilibrium chemistry on the evolution of the two-phase interstellar medium (warm and cold atomic phases), and thereby constrain the initial conditions for star formation prevailing in turbulent gas. 
We perform high-resolution simulations in three dimensions, including a realistic non-equilibrium treatment of the ionization state of the gas, and examine both 
driven and decaying turbulence. This allows us to explore variations in the metallicity $Z$. In this paper,  we
study solar metallicity, $Z=Z_\odot$, and low metallicity, $Z=10^{-3}\;Z_\odot$, gas.   
For driven, large-scale turbulence, we find that the influence of the metallicity on the amount of mass in the cold gas component is small. However, in decaying turbulent conditions this picture is much changed. 
While cold regions survive in the case of solar metallicity, they are quickly heated and dispersed in low-metallicity gas. 
This result suggests that star formation can be suppressed in environments of low metallicity, unless a strong turbulent driver is acting on time scales
shorter than a few turbulent crossing times. Inter alia this finding could explain the overall inefficient star formation as well as the burst-like mode of star formation found in metal-poor, gas-rich systems like dwarf galaxies.

\end{abstract}




\newpage
\section{Introduction}

The cold interstellar medium (ISM) and, in particular, the molecular part of it, forms the reservoir available for star formation \citep{WongBlitz2002, Kennicutt2007, Bigiel2008}. The properties of this cold gas component regulate the mass distribution of prestellar cores in a molecular cloud \citep[e.g.,][]{Padoan2002, Padoan2009, Hennebelle2008,Heitsch2008},
and hence the stellar initial mass function \citep[IMF; e.g.,][]{Motte1998, Nutter2007}. It has been suggested that  star formation within a molecular cloud is regulated by the interplay of gravity and turbulence  \citep[e.g.,][]{Klessen2000}. However, on galactic scales, understanding the formation and the properties of the cold gas reservoir available for star formation is equally important. In addition to turbulence and gravitational collapse, heating and cooling processes essentially determine the distribution of the cold ISM and therefore the global star formation efficiency \citep{Ostriker2010}.

The ISM is highly turbulent, with gas velocity dispersions of up to 
50 times the sound speed of the molecular (10 K) gas component in local galaxies, and even higher factors in high-redshift disk galaxies \citep{Genzel2006}. However, turbulence has been shown to dissipate quickly \citep{MacLow1999} if the external driving is stopped, resulting in the need to continuously drive turbulence either \textit{locally} via stellar feedback \citep{Ostriker2010} or \textit{globally} via, e.g., galaxy mergers \citep{Teyssier2010, Karl2010}, clumpy infall of cold gas \citep{Elmegreen2010, Aumer2010}, or spiral density waves. 

At the same time, the ISM is regulated by metallicity-dependent heating and cooling processes that tend to segregate it into distinct, thermally stable phases (warm and cold), which can coexist in approximate pressure equilibrium. 
The transition between the phases is driven by thermal instability \citep[TI;][]{Field1965, Burkert2000},
more specifically by its isobaric mode \citep[e.g.,][]{Pikelner1968, McKee1977, Wolfire1995}. Thus, for several reasons the efficiency with which a galaxy is able to form cold gas depends on the metal content of the gas.
First, the turbulent shock compression of the ISM depends on the metallicity-dependent cooling rate \citep[e.g.,][]{Omukai2005}. Second, within the cold component, the dust abundance (i.e., metallicity) controls the conditions under which the transition from cold atomic to molecular gas occurs \citep{Gnedin2009, Krumholz2009,Glover2010b}. Third, depending on the strength of the interstellar radiation field, low gas metallicities ($\leq 0.01 Z_\odot$) alter the balance of heating and cooling in such a way that the development of a bi-stable phase can be suppressed \citep[see semi-analytical estimates by ][]{SpaansC1997, Spaans1997}. This conclusion has also been drawn by \citet{Schaye2004} who used equilibrium models and the {\sc Cloudy} software to demonstrate that the formation of cold gas is a metallicity dependent process. His models are successfully employed to predict the outer edges of standard star forming galactic disks, for both regular spirals and dwarfs \citep{Leroy2008}. 

%%%%%
The idea that the (extra-)galactic star formation efficiencies are connected to the present ISM properties has been further discussed in  several recently developed theoretical models  \citep{Elmegreen2002, Krumholz2005, Schaye2008, Gnedin2009, Krumholz2009, Pelupessy2009, Krumholz2011}. Some of them estimate the effects of turbulence and have shown that the thermodynamics of the ISM can control the resulting shape of the density probability density function (PDF) of the ISM gas \citep{Wada2007, Tasker2008, Robertson2008}.
The numerical simulations we present here indicate that the density PDF is only strongly affected in the non-driven turbulent case.

Previous highly resolved studies of the statistics of driven or decaying turbulence mostly assumed the gas to be isothermal
\citep[e.g., ][including magnetic fields]{Kritsuk2007, Beresnyak2009}.
First steps toward a detailed numerical model of turbulence including gas thermodynamics have been published by \citet{Kritsuk2004}, \citet{Joung2006}, \citet{Brandenburg2007}, and \citet{Avillez2007}, who include parameterized cooling functions at solar metallicity.  \citet{Kritsuk2004} and \citet{Brandenburg2007} study the amount of turbulence generated by the TI itself, whereas \citet{Joung2006} and \citet{Avillez2007} use supernova explosions as internal drivers of turbulence. Other authors have studied molecular cloud formation in colliding flows in two dimensions \citep{Audit2005,Heitsch2006, Heitsch2008} and three dimensions \citep{Slyz2005, Vazquez2007, Hennebelle2008b, Banerjee2009} using parameterized cooling and heating processes. In these simulations turbulence is generated via various instabilities occurring in the collision layer of the flow.
These studies confirm the development of a multi-peaked density and temperature PDF when turbulence and equilibrium-based gas thermodynamics are treated simultaneously.
On the other hand, \citet{Vazquez2009} conclude that turbulence acts faster than TI, and thus relegates classic instabilities \citep{Field1965} to a secondary role in controlling the gas dynamics. He shows that shock-compressed diffuse gas can stay in an unstable regime for a few megayears before cooling efficiently. This is supported by recent models of \citet{Avillez2010} who point out that the treatment of non-equilibrium ionization is essential for a correct treatment of the highly dynamical ISM evolution, at least in the high temperature regime.

%%%%%%%%

In this paper, we show that the formation and evolution of the cold ISM phase is a highly metallicity dependent process. We base this conclusion on high-resolution, three-dimensional simulations with the Eulerian grid code \textsc{FLASH} \citep{Fryxell2000}, including both driven and decaying turbulence, as well as metallicity-dependent heating and cooling \citep{Glover2010}. We demonstrate the major influence of the gas metallicity in decaying conditions -- a finding which highlights the global metallicity dependence of star formation. We suggest that this result may explain puzzling observations, like the low star formation rates in metal-poor, gas-rich galaxies.

%%% IF LETTER => NO OUTLINE
%The plan of this paper is the following. 
%In section \ref{2} we describe the numerical model including gas chemistry and turbulent stirring.
%In section \ref{3} we show the development of the two-phase ISM in driven and decaying conditions,
%and we explain the differences in section \ref{4}. In section \ref{5} we draw conclusions and discuss 
%some astrophysical environments that might be affected by our findings.

%%%%%%%%%%%%% SECTION 2 NUMERICAL METHOD %%%%%%%%%%%%%%%%%%
\section{Numerical Method}\label{2}

We use FLASH 3.2, an MPI parallel Eulerian grid code, which solves the Euler equations for fluid flow. We modify the energy equation to 
include heating and cooling effects (see Section \ref{chemistry}).
%FLASH is parallelized via domain decomposition under the Message Passing Interface (MPI) and scales nearly linearly 
%with the number of processors employed (see performance and scaling behaviour).
%It is a grid-based code, which optionally features adaptive mesh refinement (AMR), and solves the hydrodynamical
%equations in three dimensions. 
The implemented directionally split, finite difference scheme is based on the 
piecewise parabolic method \citep[PPM;][]{Woodward1984, Colella1984} in its direct Eulerian form.
PPM allows one to model strong shocks and discontinuities with high accuracy.
%It is a finite volume scheme, in which the physical 
%variables are represented as zone averages.
Using the method of characteristics, the nonlinear flux of quantities between zones is obtained by solving a 
Riemann problem at each zone boundary in alternating one-dimensional sweeps through the grid. 
We use the hybrid Riemann solver (HLLE) inside shocks to avoid odd-even instabilities \citep{Quirk1997}.
%PPM makes use of second-order operator splitting \citep{Strang1968} and is therefore formally accurate to second-order both in space and time. It also uses a monotonicity constraint to control  oscillations near discontinuities rather than an artificial viscosity term.
\subsection{Turbulence}
At every time step, supersonic turbulence is driven on the largest spatial scales ($1\le k \le 3$, where $k=1$ corresponds to $2 \pi/500 \textrm{pc}$ and $k=3$ corresponds to $2 \pi/ 166 \textrm{pc}$).  
The driving energy input rate per unit mass, $dE_\mathrm{in}/dt\;=\;1.04 \times 10^{-2} \textrm{erg g}^{-1} \textrm{s}^{-1}$, is chosen such that supersonic turbulence develops,  with a roughly constant root-mean-squared (rms) velocity of $50 \: {\rm km} \: {\rm s^{-1}}$. We use a mixture of solenoidal and compressive forcing with random phases and time-correlated driving forces, which leads to coherent large-scale structures. The phases of the driving modes are modulated on a turnover timescale, $\tau_\mathrm{dyn}=50\;$Myr, which is equal to five turbulent crossing times, or about 70\% of a sound crossing time for gas at the initial temperature of 5000~K. For each mode and at each time step six separate phases (real and imaginary in three dimensions) are evolved by an Ornstein-Uhlenbeck random process, which represents a velocity in Fourier space. 
%We use direct summation to transform the velocities into real space. For a small number of modes this is more cost-efficient than using the Fast Fourier Transform. 

\subsection{Chemistry and Cooling}\label{chemistry}
Gas dynamics and gas chemistry are strongly coupled and should be solved
at the same time \citep{Glover2010}. 
In order to study the metallicity dependence of the multi-phase ISM, we
include a basic and fast chemical network, based on \citet{Glover2007b},
to follow a limited amount of non-equilibrium gas chemistry, specifically 
the chemistry involved in the determination of the ionization fraction of 
the gas. When solving the energy equation we take account of radiative cooling 
due to line emission (Ly$\alpha$, ${\rm C}_\mathrm{II}$, ${\rm O}_\mathrm{I}$, and ${\rm Si}_\mathrm{II}$) and dust,  
and the chemical cooling produced by the collisional ionization and radiative
recombination of hydrogen. We account for heating due to cosmic rays, X-rays, and 
photoelectric emission from small grains and polycyclic aromatic hydrocarbons (PAHs). For the ultraviolet background
radiation field, we use the standard estimate of \citet{Draine1978}. In this study,
we neglect the effects of dust shielding. This is a reasonable approximation in the 
warm gas in our solar metallicity simulation, but may cause us to overestimate the
temperature of the dense component; a colder dense component would  in any case strengthen our
conclusions. It is a very good approximation throughout our
low-metallicity simulation, as in this case the dust extinction is very small even
in the cold clumps. For the cosmic ray ionization rate of neutral hydrogen, we adopt
a value $\zeta_{\rm H} = 3 \times 10^{-17} \: {\rm s^{-1}}$, while for the X-ray ionization and heating
rates, we use the values given in Appendix A of \citet{Wolfire1995}, assuming
a fixed absorbing column of hydrogen ${\rm N}_\mathrm{H} = 10^{20} \: {\rm cm^{-2}}$.

We do not include the formation and destruction of molecular hydrogen (H$_{2}$) 
in our current models. This may be a significant coolant in the warm, diffuse ISM 
in the Galaxy \citep{Gnedin2010}, although this remains unclear. However, it is very
unlikely to be a significant coolant in the warm gas in our low-metallicity simulation,
as its equilibrium abundance would be very small, even if gas-phase 
formation mechanisms were included \citep{Glover2003}. We would therefore expect the
inclusion of H$_{2}$ chemistry and cooling to merely heighten the differences between the results of our solar metallicity and low-metallicity runs.

The heating and cooling terms, and also the chemical source and sink
terms, are operator split from the advection and {\it pdV} work terms and are
solved implicitly using the DVODE solver \citep{Brown1989}. In cases where
the chemical or cooling timesteps are much shorter than the hydrodynamic
time step, sub-cycling is used to treat the cooling and chemistry, thereby
avoiding the need to constrain the global time step. Chemical abundances
are represented as tracer fields and are advected using the standard FLASH
infrastructure for such fields.

\subsection{Initial Conditions and Resolution}
Our initial conditions are motivated by typical ISM values.
At $t=0$, we start from a homogeneous box with a total mass of $3.99 \times 10^6\;M_\odot$. The box length is $L=500\;\textrm{pc}$, the initial gas density $\rho_0 = 2.16 \times 10^{-24} \;\textrm{g cm}^{-3}$, and the initial temperature is set to $5000\;$K. Turbulent driving with the energy input rate given in Section 2.1 results in an rms turbulent velocity of $50\;\textrm{km s}^{-1}$, and so the initial rms Mach number is
approximately 7. We show results for two different gas metallicities: solar metallicity (hereafter \textit{Z0}) and low metallicity $Z=10^{-3} Z_\odot$ (hereafter \textit{Z-3}). We use a $512^3$ grid without AMR for all of the runs presented. 
With this setup each cell has a volume of $(0.977\;\textrm{pc})^3$.

\section{Results}\label{3}
We first evolve the box by driving turbulence for one turnover time $\tau_\mathrm{dyn}= 50\;\textrm{Myr}$. Thereafter, driving is switched off, and the gas is evolved for a further turnover time, up to $t = 100 \: {\rm Myr}$. In the following sections, we discuss the most striking features and differences of these two evolutionary stages with respect to the gas metallicity.

\begin{figure*}[htbp]
\begin{center}
\includegraphics[width=140mm]{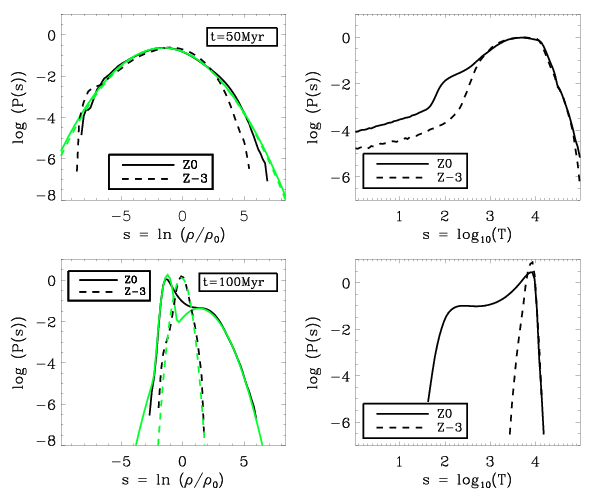}
%\begin{tabular}{r r}
%\multicolumn{2}{c}{\textbf{A:}  $t=50$Myr }\\
%\includegraphics[width=80mm]{plots/new-fig2-npdf-drv.png} &
%\includegraphics[width=80mm]{plots/new-fig2-Tpdf-drv.png}
%\end{tabular}
%\begin{tabular}{r r}
%\multicolumn{2}{c}{\textbf{B:} $t=100$Myr }\\
%\includegraphics[width=80mm]{plots/new-fig2-npdf-dec.png} &
%\includegraphics[width=80mm]{plots/new-fig2-Tpdf-dec.png}
%\end{tabular}
\caption{Density and temperature PDFs after 1 $\tau_\mathrm{dyn}$ of evolution under driven conditions (upper panels) and
after another $\tau_\mathrm{dyn}$ of evolution under decaying conditions (lower panels). Solid black lines: case \textit{Z0}; dashed black lines: case \textit{Z-3}; green lines: log-normal fits (see Table 1).}
\label{Figure 2}
\end{center}
\end{figure*}

%\footnote{Footnotes can be inserted like this.}
%\begin{figure}[htbp]
%\begin{center}
%\begin{tabular}{c}
%\textbf{A:}  $Z=Z_\odot$; $t=100$Myr \\
%\includegraphics[width=80mm]{plots/cut_z0_100Myr.png} \\
%\textbf{B:}  $Z=10^{-3} Z_\odot$; $t=100$Myr \\
%\includegraphics[width=80mm]{plots/cut_z-3_100Myr.png}
%\end{tabular}
%\caption{Like Fig. \ref{Figure 1} but evolved for one crossing time under decaying conditions. Again, panel A depicts Z0 and panel B shows Z-3.}
%\label{Figure 3}
%\end{center}
%\end{figure}

\subsection{Driven turbulence} \label{sec3_driven}
%%%%%%%% FINAL
Starting from a homogeneous gas density distribution, a two-phase medium is quickly generated due to the influence of driven, supersonic turbulence, which produces strong non-linear fluctuations in all thermodynamic variables. Kinetic energy is injected on large scales and cascades to increasingly smaller ones. This turbulence generates shocks that compress the warm, neutral medium into dense, cold filaments.

Regardless of the gas metallicity, regions with very low density ($\rho<10^{-27} \textrm{g cm}^{-3}$) and very high density ($\rho> 4\times 10^{-22} \textrm{g cm}^{-3}$) develop quickly, after only 5 Myr = 0.1 $\tau_\mathrm{dyn}$. The metallicity influences the maximum density reached, $\rho_\mathrm{max}$, 
which is about a factor of eight larger in the solar metallicity run compared to the low-metallicity run. However, this is a relatively small difference
compared to the total range of densities.

In isothermal compressible turbulence the volume-weighted density PDF is expected to follow a log-normal distribution \citep{MacLow1999, Ostriker2001, Kritsuk2007, Federrath2008, Kitsionas2009} of the form
\begin{equation}\label{logn}
p(s)ds=\frac{1}{\sqrt{2 \pi \sigma^2}} \exp{\left[-\frac{(s-s_0)^2}{2 \sigma^2}\right]}ds.  
\end{equation}  
where $s = \ln (\rho / \rho_{0})$ is the logarithmic density contrast, with $\rho_{0}$ equal to the mean gas density. 
In Equation (\ref{logn}),  $\sigma$ is the logarithmic density dispersion, and $s_0$ is the mean value of $s$. 
%Eq. \ref{logn} might be multiplied with a constant offset without changing the shape of the distribution.\\

In Figure \ref{Figure 2} (upper left panel) we show the resulting density PDFs after driving turbulence for a turnover time (i.e., five turbulent crossing times). Where appropriate, we include fits according to Equation (\ref{logn}), and summarize the resulting values for $\sigma$ and $s_0$ (or $\rho(s_0)=\rho_0 \exp{(s_0)}$, respectively) in Table 1. In the driven case, the density PDFs are well described with a single log-normal function for both \textit{Z0} and \textit{Z-3}. For \textit{Z0} the distribution is slightly asymmetric and skewed toward higher densities. However, the best fits result in a similar variance of $\sigma \simeq 1.7$ in both cases. Studies of driven turbulence in isothermal gas have found a relationship between $\sigma$ and the rms Mach number $M$ \citep{Padoan1997,Padoan2002, Price2011}
\begin{equation}
\sigma^{2} = \ln \left(1 + b^{2} M^{2} \right),  \label{pn}
\end{equation}
where the value of the parameter $b$ depends on the relative importance of the compressive and solenoidal modes \citep{Federrath2008}. For the driving used in our simulations, $b = 0.5$, and the rms Mach number at $t = 50 \: {\rm Myr}$ is 8.66 in the solar metallicity simulation, and 9.26 in the low-metallicity simulation. Equation (\ref{pn}) therefore predicts that $\sigma = 1.72$ for the solar metallicity simulation and $\sigma = 1.76$ for the low metallicity simulation, in good agreement with our measured value of 1.7.

We emphasize that our finding of a broad, but single-peaked density PDF supports the conclusions of \citet{Vazquez2009}, where he argues that supersonic turbulence should dominate the evolutionary state of ISM gas because it acts faster than conventional TIs. We note that a single-peaked density PDF has also been found by \citet{Avillez2007}. However, models by e.g., \citet{Slyz2005}, who drive turbulence in colliding flows, result in a double-peaked density PDF. We suggest that the transition between single- and double-peaked density PDF profile depends on the strength of the turbulence, i.e., the Mach number. This suggestion is verified below (Section 3.2).\\

The temperature PDFs, which are also shown in Figure \ref{Figure 2} (upper right panel) cannot be well approximated with a single log-normal distribution, which only gives acceptable fits for the high temperature parts. Toward low temperatures (i.e., for $T<300$K), the distributions show extended powerlaw tails. This behavior is caused by the cooling instability that separates the warm and cold neutral gas. \\

Overall, we find the temperature distributions to be more strongly affected by the gas metallicity than the density PDFs. In particular, the volume filling factors (VFFs) of the cold components differ significantly. For further analysis, we define \textit{the cold component} to be all gas with $T<300$K, as 300 K is the temperature threshold below which a stable, cold atomic phase can exist \citep{Wolfire2003}.
Then we find 1.7\% of the computational domain to be cold for \textit{Z0}, whereas we find only 0.075\% to be cold for \textit{Z-3}.\\

\begin{figure*}[htbp]
\begin{center}
\begin{tabular}{@{\hspace{2.7cm}}c @{\hspace{3.3cm}}c @{\hspace{3.3cm}}c c}
\textbf{A} & \textbf{B} & \textbf{C} & \textbf{D} \\
\multicolumn{4}{r}{\includegraphics[width=160mm]{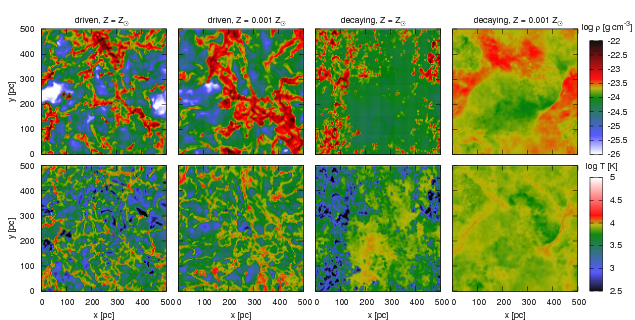}}
\end{tabular}
\caption{Cuts through the computational volume on the $(z=0)$-plane, showing the density and temperature of the developed two-phase media after supersonic turbulence has been driven for five turbulent crossing times (panels (A) and (B)), and then allowed to decay for a further five turbulent crossing times (panels (C) and (D)). Panel (A) and (C) depict the solar metallicity case, \textit{Z0}, whereas panel (B) and (D) show the low-metallicity case, \textit{Z-3}.}
\label{Figure 1}
\end{center}
\end{figure*}

In Figure \ref{Figure 1}, columns A and B,  we show the density (top panels) and temperature (bottom panels) of the evolved filamentary structures in the $z=0$ plane. Prima facie the density distributions of the solar metallicity (panel (A)) and the low-metallicity case (panel (B)) look similar, in agreement with the PDFs. However, the filamentary structure formed in run \textit{Z-3} seems to be coarser and individual filaments seem to be less condensed than in run \textit{Z0}. \\

When comparing the two-dimensional density and temperature distributions (Figure \ref{Figure 1}), we find both shock heating and {\it pdV} heating as well as radiative cooling and cooling by expansion to be very important. The presence of shock-heated regions surrounding dense and cool filaments can be found in both the \textit{Z0} and the \textit{Z-3} case. However, most of the cold regions in run \textit{Z-3} can be identified with voids in the density structures (see also Figure 3). Thus, these regions have cooled due to expansion. Radiatively cooled regions within dense filaments are very rare in run \textit{Z-3}, whereas these dominate the cold part in run \textit{Z0}. It should be noted that for \textit{Z0} the mean density in the cold gas is higher than the mean density in the warm gas ($T>300$K), as expected for gas which truly undergoes a cooling instability. However, for \textit{Z-3} this is not the case (see Table 1).
This indicates that cooling due to expansion is dominant in case \textit{Z-3} whereas the dense, cold component is very small. Thus, our results confirm previous semi-analytical estimates \citep[e.g., ][]{Schaye2004} that the formation of stable molecular cloud cores might be strongly inhibited in low-metallicity gas.\\

%%%%%%%%%%%%%%%%%%%%%%%%%%%%%%%%%%%%%%%%%

\begin{table*} \label{table 1}
\begin{center}

\begin{tabular}{lcccc | c c}
\tableline\tableline
Run & Time & Max & Mean Warm & Mean Cold & 
\multicolumn{2}{c}{Fit} \\
 &$[\textrm{Myr}]$& $[\textrm{g cm}^{-3}]$& $[\textrm{g cm}^{-3}]$& $[\textrm{g cm}^{-3}]$&
 $\rho(s_0)$ & $\sigma$ \\
\tableline
& & & & & \multicolumn{2}{c}{}\\
\textit{Z0}& 50 & $2.4 \times 10^{-21}$ & $1.65 \times 10^{-24}$ & $3.35 \times 10^{-23}$ & $4.4 \times 10^{-25}$ & 1.73\\
\textit{Z-3}& 50 & $3.2 \times 10^{-22}$ & $2.07 \times 10^{-24}$ & $7.33 \times 10^{-25}$ & $4.6 \times 10^{-25}$ & 1.70\\
\textit{Z0}& 100 & $1.0 \times 10^{-21}$ & $5.83 \times 10^{-25}$ & $1.22 \times 10^{-23}$ & $6.3 \times 10^{-25}$ & 0.22 \\
& & & & & $8.2 \times 10^{-24}$ & 0.93 \\

\textit{Z-3}& 100 & $ 1.2 \times 10^{-23}$ & $ 9.27 \times 10^{-25}$  & -- & $1.9 \times 10^{-24}$ & 0.31 \\
%& & & & & \multicolumn{3}{c}{}\\
\tableline \tableline
\end{tabular}
\caption{Absolute Maximum (Column 3) and Mean Densities (Columns 4 and 5) Extracted from the Simulations. The first two rows are for the part of driven turbulence, the following three rows are for the part of decaying turbulence. The mean densities are time averages over 50 Myr for the warm ($T \ge 300$K, Column 4) and cold ($T < 300$ K, Column 5) gas components individually. For reference, the mean gas density summed over both components was $2.16 \times 10^{-24} \;\textrm{g cm}^{-3}$ throughout. We also list the parameters for the log-normal fits to the density PDFs shown in Figure \ref{Figure 2}. The parameters are described in Equation (\ref{logn}). For run \textit{Z0} we list two sets of parameters at $t=100$ Myr as this distribution can only be fitted with the sum of two log-normal functions (see Figure 1).} 
\end{center}
\end{table*}

%%%%%%%%%%%%%%%%%%%%%%%%%%%%%%%%%%%%%%%%%%%%

\subsection{Decaying turbulence}\label{sec3_decay}

Starting from the evolved turbulent setup discussed before, we switch off turbulent driving at $\tau_\mathrm{dyn}=50$ Myr, and study the evolution of the gas under decaying conditions until $t=2\;\tau_\mathrm{dyn}$. 
In Figure 3 (upper panel) we show the evolution of the rms velocity dispersion, $v_\mathrm{rms}$, for the full time interval. After driving has been switched off $v_\mathrm{rms}$ decreases. From initially close to $50 {\rm km s}^{-1}$, $v_\mathrm{rms}$ drops down to $20 {\rm km s}^{-1}$ within 5 Myr ($t = 55$ Myr) and to $10 {\rm km s}^{-1}$ within approximately 25 Myr ($t = 75$ Myr). In Figure 3 (lower panel) we depict the decay of kinetic energy. We find that both cases, \textit{Z0} and \textit{Z-3}, are well described by the fitting formula derived by \citet{Pavlovski2002}
\begin{equation}
E(t) = E_0 (1+t/t_1)^\eta,
\end{equation}
where $E_0$ is the initial kinetic energy and $t_1= L_{\rm inj}/v_\mathrm{rms}$ is equal to the initial ratio of mean energy injection scale to the rms velocity. For isothermal, compressible MHD turbulence, \citet{MacLow1999} derived $\eta \approx -1$. \citet{Kitsionas2009} also extracted $\eta = -1$ for compressible HD turbulence. Incompressible, non-magnetic turbulence has been found to decay somewhat faster with $\eta$ in $[-1.2, -1.4]$. In addition, cooling has been found to accelerate the decay as the average Mach number is larger \citep{Kritsuk2002, Pavlovski2002}. With cooling and MHD turbulence, \citet{Pavlovski2002} derived $\eta \approx -1.3$, whereas \citet{Kritsuk2002} found indices between $-1$ and $-2$. 
We derive $\eta \approx -1.3$ for \textit{Z-3}. For \textit{Z0}  we find a slow decay within the first 10 Myr ($\eta \approx -1$), but a steep decay thereafter with $\eta \approx -1.8$ from 10 Myr ($t = 60$ Myr) onward. Due to the fact that cooling is much more efficient in the case of solar metallicity, {\it Z0} (see Section 4.1 for a discussion of the mass fraction in cold gas), and based on previous findings that the decay is accelerated with cooling, our findings are in overall agreement with the expectation that $\eta$ should be lower for {\it Z0} than for {\it Z-3}. \\

Returning to Figure 1, we show the temperature and density structure of run \textit{Z0} (panel (C)) and run \textit{Z-3} (panel (D)) after turbulence has decayed for 50 Myr. In this case we find striking differences between \textit{Z0} and \textit{Z-3}. Whereas cold clumps survive in run \textit{Z0}, they are completely dispersed in run \textit{Z-3}. In fact, we find all of the cold gas in \textit{Z-3} to be warmer than 300 K after 10 Myr (i.e., at $t=60$ Myr). \\

In agreement with this qualitative finding, the metallicity also has a major impact on both density and temperature PDFs (see lower panels in Figure \ref{Figure 2}). All distributions show a significant decrease in variance as the system evolves toward its equilibrium density and temperature configuration. 
In run \textit{Z-3} a single-peaked density PDF remains. However, now the variance of the distribution is so small that we can speak of a single, warm phase. All of the cold gas has been heated and therefore the VFF of the cold component is 0. In this environment star formation is basically impossible. This is in agreement with the findings of \citet{Schaye2004} and \citet{Spaans1997}, who predict that star formation is strongly suppressed in low-metallicity environments.

For run \textit{Z0}, the single log-normal density PDF is replaced by a combination of two log-normals representing two distinct gas phases, a warm diffuse and a cold dense phase. 
A double-peaked density PDF has been previously reported by authors like \citet{Slyz2005} and \citet{Audit2010}, who employed cooling functions. In contrast to their work, the separation into two distinct phases seems to be less clean in our case. 
We note that the blurring of the warm and cold phases is probably a consequence of our chemo-dynamical model. We expect the separation to become increasingly clear as the turbulence decays further and $v_\mathrm{rms}$ decreases. We will discuss the time evolution of the density PDFs for different turbulent driving energies in a subsequent paper.
Finally, the VFF or of the cold component is slowly increasing in the case of {\it Z0}, and reaches 4\% at $t=100$ Myr. 

\begin{figure}[htbp]
\begin{center}
\includegraphics[width=80mm]{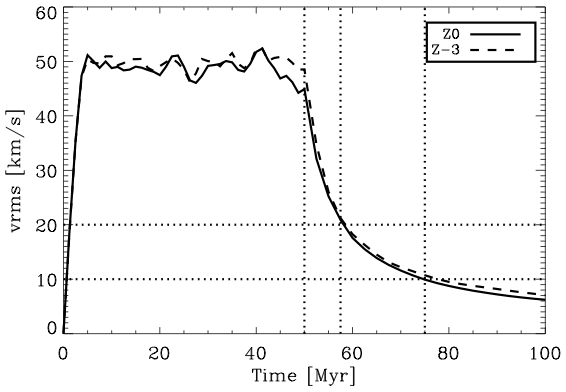}
\includegraphics[width=80mm]{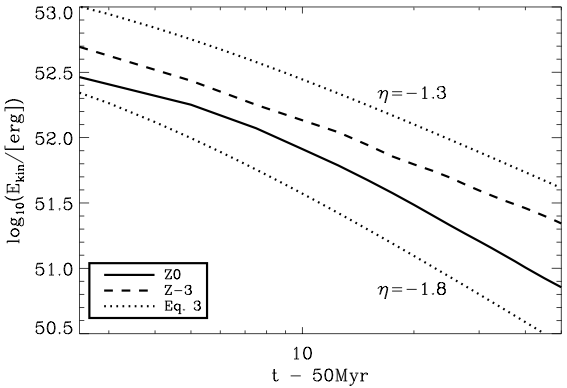}
\caption{Upper panel: we depict the root mean square velocity as a function of time. From left to right, the first dotted vertical line marks the beginning of the decay phase at $t=50$ Myr, the following two dotted horizontal and vertical lines mark the points at which the turbulence has decayed to $v_\mathrm{rms} = 20 {\rm km s}^{-1}$ (55 Myr) and  $v_\mathrm{rms} = 10 {\rm km s}^{-1}$ (75 Myr), respectively.  Lower panel: decay of kinetic energy for the second 50 Myr of evolution. We show both cases, \textit{Z0} (solid line) and \textit{Z-3} (dashed line), as well as the decay curves derived from Equation (3) (dotted lines) for $\eta=-1.3$ and $\eta=-1.8$. The kinetic energy decay is faster in solar metallicity gas.}
\label{Fig_decay}
\end{center}
\end{figure}

%%%%%%%%%%%%%%%%%%%%%%%%%%%%%%%%%%%%%%%%%%%%%%%%%%%
%% The \notetoeditor{TEXT} command allows the author to communicate
%%%%%%   \notetoeditor{Figures 1 and 2 should appear side-by-side in print} 
\section{Discussion}\label{4}

\begin{figure*}[htbp]
\begin{center}
\includegraphics[width=80mm]{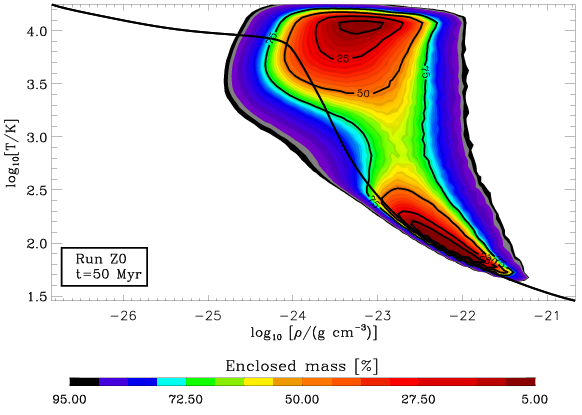}
\includegraphics[width=80mm]{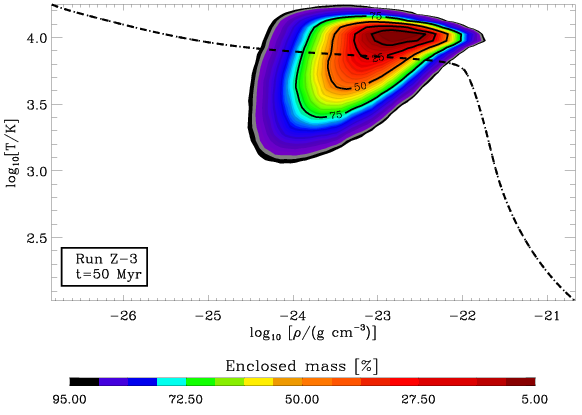}
\includegraphics[width=80mm]{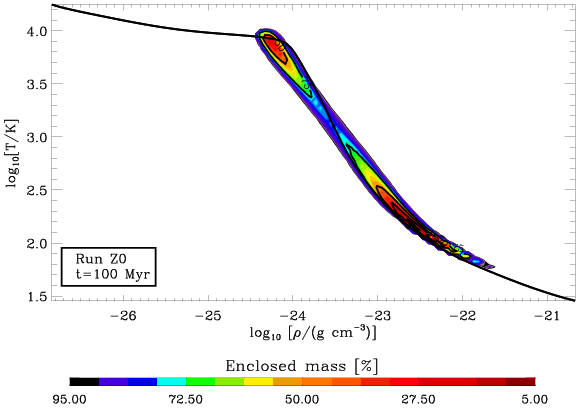}
\includegraphics[width=80mm]{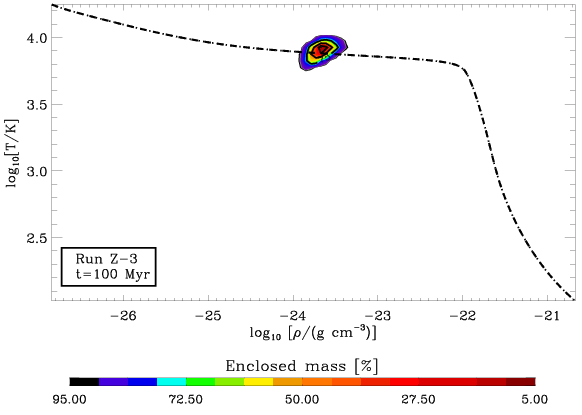}
\includegraphics[width=80mm]{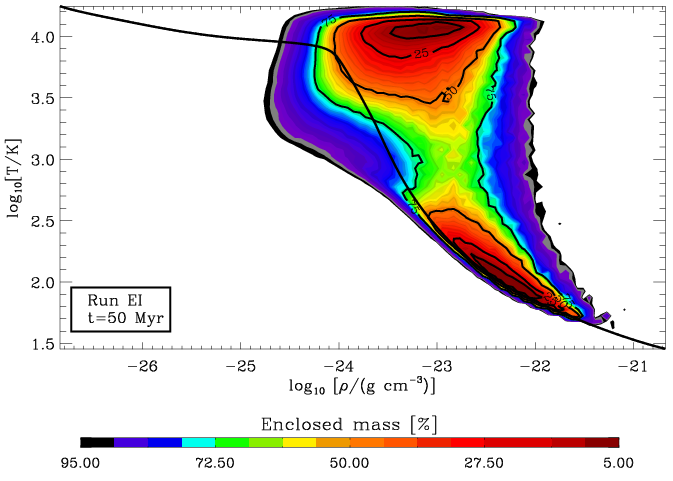}
\includegraphics[width=80mm]{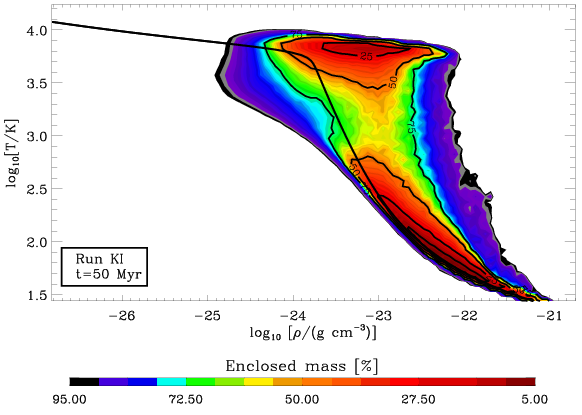}

\caption{Cumulative mass distributions derived for driven turbulence at $t=50$ Myr (top row; see Figure 2 columns A and B) and for decaying turbulence at $t=100$ Myr (middle row; see Figure 2 columns C and D). The equilibrium curves of density and temperature for $Z_0 = 1 \: Z_{\odot}$ (black solid line, left column) and $Z_0=10^{-3} \: Z_{\odot}$ (black dashed line, right column) are plotted on top. For comparison we show the mass distributions of runs {\it EI} and {\it KI} (bottom row), both at the end of the driven period at $t=50$ Myr. See Section 4.2 for further details on these runs {\it EI} and {\it KI}. }
\label{Figure 4}
\end{center}
\end{figure*}

\begin{figure}[htbp]
\begin{center}
\includegraphics[width=85mm]{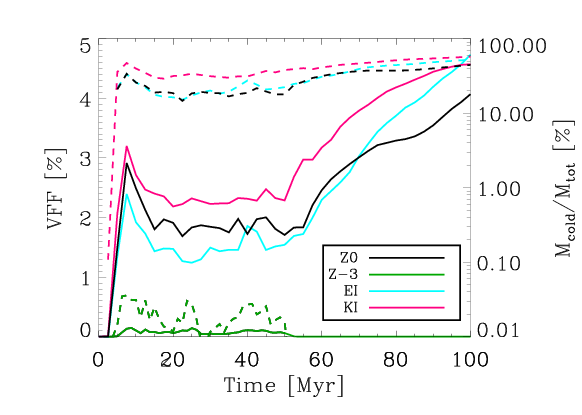}
\caption{Time evolution of the volume filling factor (solid lines) and the mass fraction (dashed lines) of the cold gas ($T<300$ K). Turbulent driving was switched off at $t=50$ Myr. We show all four runs: \textit{Z0} (black), \textit{Z-3} (green), \textit{EI} (turquoise), and \textit{KI} (magenta).}
\label{Figure 5}
\end{center}
\end{figure}

In the following, we discuss the impact of turbulence and non-equilibrium chemistry on the dynamical evolution of the ISM. 

\subsection{Metallicity dependence}
 In Figure \ref{Figure 4} (upper and central panels) we show the equilibrium density-temperature curves for low and solar metallicity together with the mass distributions found in {\it Z0} and {\it Z-3} at 50 Myr and 100 Myr. At low densities ($\rho \stackrel{<}{_\sim}10^{-24} \textrm{g cm}^{-3}$) and high temperatures, cooling is dominated by Ly$\alpha$ and therefore the metallicity dependence is negligible. For higher densities and lower temperatures, the cooling rate is dominated by fine structure emission from $\textrm{C}^{+}$ and O, and is therefore strongly dependent on $Z$. \\
Based on comparing the equilibrium curves for low and high $Z$, it might at first seem surprising that both runs ({\it Z0} and {\it Z-3}) look so similar in case of driven turbulence and show a log-normal profile which is single-peaked, just like in an isothermal case. Therefore we conclude that the shape of the density PDF (single or double peaked; see Figure \ref{Figure 2}) is not only a function of the employed gas thermodynamics but at the same time must be a function of the turbulent Mach number. 
Strongly driven turbulence \citep{Vazquez2009} causes the gas to populate a large region of parameter space in the ($\rho,T$)-plane significantly displaced from the equilibrium locus. As a result, the density PDF in the case of driven turbulence is broad, but single-peaked just like in an isothermal simulation (Figure \ref{Figure 2}, upper panels). Even at late stages of the evolution ($t=100$ Myr), where $v_\mathrm{rms}$ has decayed to $7 \;{\rm km s}^{-1}$, the separation into two phases is less distinct than in previous models \citep{Slyz2005, Audit2010}. A closer investigation reveals that, apart from the different thermodynamics schemes employed, also the Mach number seems to be lower in these models than in our simulations. For instance, \citet{Audit2010} who reported a clearly double peaked profile had generally low turbulent Mach numbers (the Mach number is around 1 in \citeauthor{Audit2010} \citeyear{Audit2010}). On the other hand, authors who used efficient driving mechanisms \citep{Avillez2007} also reported single-peaked distributions despite including gas heating and cooling. We therefore infer that a pronounced double-peaked density PDF is a sign of a two-phase medium at low Mach numbers.\\

Even though the density PDFs in the driven turbulent cases look similar and seem to be independent of the gas metallicity, we find the respective mass distributions in the $(\rho,T)$-plane to be quite different. For {\it Z0}, a large fraction of the gas is in the thermally unstable regime, in agreement with observations \citep[e.g.,][]{Kanekar2003}. Moreover, for \textit{Z0}, low temperature values preferentially coincide with high-density values. For \textit{Z-3}, the opposite trend is found. Here, lower temperature values correspond to lower densities. In general, only a small mass fraction is found to be at temperatures less than 1000 K, as cold gas preferentially coincides with regions of low density, where expansion is the main cooling process.

We now determine the fraction and the VFF of the cold gas formed in our simulations.
 For decreasing metallicity the equilibrium temperature, $T_\mathrm{equi}$, remains high ($T_\mathrm{equi}>$ 300 K) up to increasingly high density. For further reference, we define the threshold density $\rho_\mathrm{thres}$ above which gas may cool below 300 K. For $Z_0=Z_\odot$, we find $\rho_\mathrm{thres} \simeq 9 \times 10^{-24}\;\textrm{g cm}^{-3}$, and for $Z_0=10^{-3} Z_\odot$ we find $\rho_\mathrm{thres} \simeq 5 \times 10^{-22}\;\textrm{g cm}^{-3}$. Thus, for $Z_0=Z_\odot$ some of the gas reaches densities higher than $\rho_\mathrm{thres}$, whereas for $Z_0=10^{-3} Z_\odot$, $\rho_\mathrm{max}(Z-3)=3.2 \times 10^{-22}  \textrm{g cm}^{-3}$ and $\rho_\mathrm{thres} > \rho_\mathrm{max}$ (see Section 3.1 and Table 1).\\
 
The strong influence of the gas metallicity on the formation and survival of cold coherent clumps only reveals itself during the non-driven, decaying phase, as the temperature and density PDFs of the two simulations are very similar during the driven turbulent period.  After turbulent driving has been switched off (Figure \ref{Figure 4}, middle panels) the systems evolve toward local equilibrium. Consequently, the almost immediate disappearance of the cold component in run \textit{Z-3} can be explained. The cold component formed in the driven case was solely a product of turbulent stirring. Regions of high density, cold gas, which truly undergo thermal instability are rare at low metallicity since the high threshold density required for the formation of a stable cold component cannot be reached. Therefore, when evolved under decaying conditions, the cold gas is quickly heated up and no cold and dense clumps remain.

This result is reinforced by the time evolution of the VFF of the cold gas component, and of the mass in cold gas, shown in Figure \ref{Figure 5}. With driven turbulence the VFF is roughly constant and about 20 times lower in \textit{Z-3} than in \textit{Z0}. With decaying turbulence the two cases diverge as VFF is slowly increasing for \textit{Z0} whereas it quickly approaches 0 for \textit{Z-3}. The mass fraction of gas found in the cold component evolves in a similar manner as the VFF. However, even in the driven case, the mass in cold gas is more than a factor of 1000 lower for \textit{Z-3} than for \textit{Z0} (with an average of $10^3 \;M_\odot$ for \textit{Z-3}, corresponding to only 0.025\% of the total mass, compared to an average of $10^6 \;M_\odot$ for \textit{Z0}, corresponding to a substantial fraction of the total mass, i.e., 25\%).

%%%%%%%%%%%%%%%%%%%%%%%

\subsection{The impact of non-equilibrium chemistry}
\begin{figure}[htbp]
\begin{center}
\includegraphics[width=80mm]{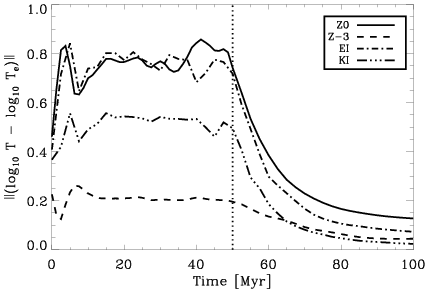}
\caption{Mean deviation of actual gas temperature and equilibrium temperature as a function of time, again for {\it Z0} (solid line) and {\it Z-3} (dashed line). In addition, we show run {\it EI} (dash-dotted line) and run {\it KI} (dash-dot-dot-dotted line).}
\label{Fig_ttequi}
\end{center}
\end{figure}

In the following, we analyze the impact of non-equilibrium chemistry on our findings. To quantify the degree of any departures from thermal equilibrium in runs {\it Z0} and {\it Z-3} we first monitor the mean difference between actual gas temperature and expected equilibrium temperature in Figure \ref{Fig_ttequi} (solid and dashed line). In the case of solar metallicity, {\it Z0}, we find a significant offset, $\Delta T$, of the actual gas temperature from the equilibrium temperature, in agreement with Figure 4 (upper left panel). In the decaying period ($t > 50$Myr) $\Delta T$ then decreases, as does the turbulent Mach number. For {\it Z-3}, the system is close to equilibrium throughout the simulation and $\Delta T$ is small.

Two possible causes of the high $\Delta T$ in run {\it Z0} are: (1) non-equilibrium chemical effects and (2) efficient thermalization of turbulent energy on a time scale shorter than the cooling time scale. To investigate the importance of the non-equilibrium chemistry on the gas heating and cooling, we performed two comparison runs at $Z=Z_\odot$ that do not take into account non-equilibrium effects. In run {\it EI}, we still consider the detailed Glover et al. cooling function when solving the energy equation, but at the same time we assume chemical equilibrium. In run {\it KI} we completely neglect the chemical network as well as the detailed cooling function, and instead use the parameterized cooling function:
\begin{eqnarray}
\Gamma & = & 2.0 \times 10^{-26} \textrm{erg\; s}^{-1}, \\
\frac{\Lambda(T)}{\Gamma} & = &10^7 \exp{\left(\frac{-1.184 \times 10^5}{T+1000}\right)} \nonumber \\
 & + &1.4 \times 10^{-2}  \sqrt{T} \exp{\left(\frac{-92}{T}\right)} \;\;\textrm{cm}^{3},\;\;\;\;\;
\end{eqnarray}
which is based on \citet{Koyama2000, Koyama2002} \citep[see also][]{Vazquez2007}.

We include the results for both {\it EI} and {\it KI} in Figures \ref{Figure 4} (bottom row), \ref{Figure 5}, and \ref{Fig_ttequi}.
From Figure \ref{Figure 4}, we find that the phase space mass distribution formed in run {\it EI} is almost identical to run {\it Z0}. Thus, non-equilibrium chemistry seems to have a minor effect at high driving velocities. We conclude that the offset of the mass distribution with respect to the equilibrium curve as well as the high $\Delta T$ in Figure \ref{Fig_ttequi} are caused by turbulent shock compression and extra heating associated with turbulent dissipation. Run {\it KI} supports this picture, although we find {\it KI} to differ from  {\it EI} and {\it Z0} in mass distribution (see Fig. \ref{Figure 4}, bottom right panel) and mean temperature (see Figure \ref{Fig_ttequi}). Overall, run {\it KI} is closer to thermodynamical equilibrium throughout the whole simulated time.

In order to elaborate the global effect of using a detailed cooling function together with a chemical network rather than a simple, parameterized cooling function, we calculated the time evolution of the cold mass fractions and the VFFs for {\it EI} and {\it KI} and included them in Figure \ref{Figure 5} (turquoise and magenta lines). We find that run {\it EI} is very similar to {\it Z0} throughout the driven phase, although cooling seems to be slightly stronger in case of {\it Z0}, as the VFF of the cold component is somewhat higher with respect to {\it EI}. On the other hand, in the decaying phase, this effect is reversed after $t = 70$ Myr, where $v_\mathrm{rms} \approx 13 {\rm km s}^{-1}$. The reduced cooling occurring in run {\it Z0} with full chemistry is also visible in the cold gas fraction, which drops below {\it EI} after $t = 70$ Myr. \\

In summary, we cannot see strong immediate effects of the non-equilibrium chemistry.  We note that this finding might have to be revised once the formation of molecules (e.g., ${\rm H}_2$) is considered \citep{Gnedin2011}. 
Furthermore, our simulations show that using a detailed, time-dependent cooling prescription is still superior to using a parameterized cooling function as the results are much more accurate. Run {\it KI} continuously overpredicts the fraction of cold gas  (by up to 200\%) and the VFF of the cold gas.\\

%%%%%%%%%%%%%%%%%%%%%
\subsection{Other effects that may influence the phase distribution}

As the variance of the density PDF has been shown to be directly related to the turbulent Mach number \citep{Padoan1997,Padoan2002,Federrath2008}, the spread in $\rho$ about the given equilibrium configuration is expected to increase with increasing Mach number $M$. More compressive driving would enhance this effect \citep{Federrath2010}. Very strong turbulent driving could therefore result in the formation of a stable cold phase even if the metallicity were low. Also our models do not include the effects of  self-gravity. Self-gravity is known to cause a power-law tail toward high densities in the density PDF \citep{Scalo1998, Klessen2000, Wada2001, Li2003, Slyz2005, Kritsuk2011, Cho2011} and therefore would certainly influence the gas density distribution at high $\rho$. As the powerlaw tail is typically flatter than the log-normal distribution (with power spectral indices of around $-1.5$) more dense gas would be available for cooling, and the cold component would become more massive leading to more star formation. Therefore, the effect of self-gravity might shift the critical values of turbulent Mach number above which the formation of cold gas is suppressed, and the critical values of metallicity below which the formation of cold gas is suppressed.\\

The assumed FUV heating rate, $\Gamma_\mathrm{FUV}$, also influences the distribution of phases. In our present study, we assume that the photoelectric heating rate per unit dust mass is the same in both simulations, so that the only change in $\Gamma_\mathrm{FUV}$ between the two simulations comes from the change in the total dust mass, which we assume to scale linearly with metallicity. In reality, the composition and size distribution of dust grains in very low metallicity gas may be quite different from what we are used to in the Milky Way, which would lead to an additional change in $\Gamma_\mathrm{FUV}$. The strength of the FUV radiation field is also an influential uncertainty. Decreasing the FUV intensity by a factor of 100 results in a decrease in $\rho_\mathrm{thres}$ of about a factor of 10 \citep{Wolfire2003}, thus facilitating the formation of a cold component in regions of low star formation. In an upcoming paper we will further discuss the relationship between $\rho_\mathrm{thres}$, $Z$, and $\Gamma_\mathrm{FUV}$, as well as that between VFF and $M$.

Placed in an astrophysical context, our finding implies that star formation would effectively be inhibited in a system of low metallicity if no event capable of driving strong turbulence (e.g., spiral shocks in a disk galaxy, a merger, or the infall of gas onto a galactic disk) occurred.

%%%%%%%%%%%%%%%%%%%%%%%%%%%%%%%%%%%%%%%%%%%%%%%%%%

\section{Conclusions and Astrophysical Context}\label{5}

The dependence of the cooling rate on the metal abundance in the gas shifts the pressure range that allows for the formation of a bi-stable medium toward higher values with decreasing metallicity. 
Even in the Milky Way, this effect is so influential that the formation of cold molecular gas is prevented at distances larger than 18 kpc from the Galactic Center where $Z\approx 0.2\;Z_\odot$ \citep[see analytical estimates by ][]{Wolfire2003, Elmegreen1984}.

In order to understand this process in greater detail, we have simulated the turbulent evolution of a multi-phase ISM under driven and decaying turbulent conditions at different metallicities ($Z=Z_\odot$ and $Z=10^{-3} Z_\odot$). This mimics interstellar conditions for different galactic environments.
For driven turbulent conditions, we find that metallicity does not have a strong influence on either the shape of the resulting, single-peaked density PDF, or the temperature PDF. However, for $Z=10^{-3} Z_\odot$ the VFF of the cold component is lower by an approximately constant factor of 20.

In decaying turbulent conditions, we find a clear relation between the abundance of cold gas and the metallicity. Whereas the coldest regions are quickly dispersed in case of low metallicity and the system evolves toward a single component, warm medium, a stable cold component is formed in case of solar metallicity and a double-peaked density PDF is developed. Thus, on a dynamical time scale, the survival of cold clumps is strongly influenced by the gas chemistry. In agreement with previous models of other authors \citep[e.g.,][]{Schaye2004}, our numerical simulations imply that the turbulent evolution of the ISM at different metallicities causes systematic changes in the initial conditions for star formation and should strongly affect the global star formation efficiency of a galaxy. \\

We also find the non-equilibrium chemistry to not have a significant influence on our results at high turbulent velocity dispersions for both, low and solar metallicity. However, in mildly turbulent conditions ($v_\mathrm{rms} \stackrel{<}{\sim} 13 \;\textrm{km s}^{-1}$), the effect of non-equilibrium chemistry becomes more pronounced. Moreover, we find that using an accurate cooling prescription, like the one presented in \citet{Glover2007b}, rather than a parameterized cooling function is especially important to precisely predict the mass and VFF of the cold gas component in a multi-phase model.\\

Placed in a more global astrophysical context these results could be meaningful to several fields. We can speculate on the following.
\begin{itemize}
\item Metal-poor dwarf galaxies have very low star formation rates \citep{Boissier2003}, which can currently not be reproduced in simulations of dwarf galaxy formation and evolution \citep{Guo2010, Gnedin2010}. If an external trigger (e.g., a merger or tidal perturbation) were missing, our chemo-dynamical ISM model would lead to such a low star formation rate. 

\item Dwarf galaxies seem to have complex star formation histories \citep[e.g., as observed for the Carina dSph by][]{Tolstoy2003, Burkert1997}. In our model, metal-poor dwarfs could only form stars if the gas were stirred. Hence, clumpy gas infall, tidal effects, or other forms of irregular stirring (e.g., due to gravitational instabilities) would naturally lead to very complex star formation histories in these systems. 

\item Metal-poor galaxies which have a low star formation rate should be able to slowly accrete a lot of gas -- e.g., via gas infall in cold streams \citep{Keres2005, Dekel2009} -- without experiencing a significant increase in their star formation rate. Therefore, they would acquire the potential to become progenitors for massive starburst galaxies. 

\item The star formation rate in merging galaxies is probably not strongly dependent on the gas metallicity if supersonic gas turbulence is driven due to the merger.
\end{itemize}


%% Included in this acknowledgments section are examples of the
%% AASTeX hypertext markup commands. Use \url without the optional [HREF]
%% argument when you want to print the url directly in the text. Otherwise,
%% use either \url or \anchor, with the HREF as the first argument and the
%% text to be printed in the second.

\acknowledgments
We thank the anonymous referee for his/her suggestions, which significantly improved the manuscript. SW thanks Dr. Thorsten Naab for the useful discussions. S.W. and A.W. acknowledge the support of the Marie Curie RTN \textsc{Constellation} (MRTN-CT-2006-035890). SW especially thanks the neuro-surgical team -- Prof. Trappe, Dr. Weinzierl, Dr. Jaeger, and Dr. Egermann -- at Klinikum Freising for their excellent work and the inspiring environment.
R.W. acknowledges support from the Institutional Research Plan AV0Z10030501 of
the Academy of Sciences of the Czech Republic and project LC06014--Centre for
Theoretical Astrophysics of the Ministry of Education, Youth and Sports of the
Czech Republic.
S.G. acknowledges support from a Heidelberg University Frontier Grant, funded as part of the German Excellence Initiative,
from the Bundesministerium f\"ur Bildung und Forschung via the ASTRONET project STAR FORMAT (grant  05A09VHA),
and from the Landesstiftung Baden-W\"urttemberg via their program International Collaboration II (grant PLS-SPII/18).
AW further acknowledges the support of Grant ST/HH001530/1 from the UK Science and Technology Facilities Council.
The simulations shown in this work have been performed on the ARCCA SRIF-3 cluster \textsc{Merlin} in Cardiff.
The software used in this work was in part developed by the DOE-supported ASC/Alliance Center for Astrophysical Thermonuclear Flashes at the University of Chicago. 
%For technical support, please write to \email{aastex-help@aas.org}.




\bibliography{references.tex}

\clearpage

\end{document}